\documentstyle[prl,aps,psfig,floats]{revtex}
\def\etal{{\it et al.}}

\def\kmsmpc{km~s$^{-1}$~Mpc$^{-1}$}
\def\eg{{\it e.g.}}
\def\ie{{\it i.e.}}
\def\omegam{\Omega_{\rm M}}

\def\omegal{\Omega_{\phi}}
\def\omega0{\Omega_0}
\def\spose#1{\hbox to 0pt{#1\hss}}
\def\simle{\mathrel{\spose{\lower 3pt\hbox{$\mathchar"218$}}
     \raise 2.0pt\hbox{$\mathchar"13C$}}}
\def\simge{\mathrel{\spose{\lower 3pt\hbox{$\mathchar"218$}}
     \raise 2.0pt\hbox{$\mathchar"13E$}}}
\def\apj{{\it Astroph.~J.}}
\def\mn{{\it Mon.~Not. Roy.~ast. Soc.}}

\def\aj{{\it Astron.~J.}}
\def\prl{{\it Phys.~Rev. Lett.}}
\def\prd{{\it Phys.~Rev.~D}}

\def\nat{{\it Nature}}
\tighten 
\begin{document}
\draft
\preprint{IUCAA-41/99}
\twocolumn[
\hsize\textwidth\columnwidth\hsize\csname@twocolumnfalse\endcsname
\title{Reconstructing the Cosmic Equation of State
from Supernova distances}
\author{Tarun Deep Saini${}^{a}$, 
Somak Raychaudhury${}^{a}$, 
Varun Sahni${}^{a}$
and A. A. Starobinsky${}^{b,c}$}

\address{${}^{a}$ Inter-University Centre for Astronomy \& Astrophysics,
Pun\'e 411 007, India}
\address{${}^{b}$ Landau Institute for Theoretical Physics, 
117334 Moscow, Russia}
\address{${}^{c}$ MPI f\"ur Astrophysik, 86740 Garching bei M\"unchen,
Germany}
\date{\today}
\maketitle
\begin{abstract}
Observations of high-redshift supernovae indicate that the universe is
accelerating.  Here we present a {\em model-independent} method for
estimating the form of the potential $V(\phi)$ of the scalar field
driving this acceleration, and the associated equation of state
$w_\phi$.  Our method is based on a versatile analytical form for the
luminosity distance $D_L$, optimized to fit observed
distances to distant supernovae and differentiated to yield $V(\phi)$
and $w_\phi$. Our results favor $w_\phi\simeq -1$ at the present
epoch, steadily increasing with redshift.  A cosmological constant is
consistent with our results.
\end{abstract} 

% for PACS codes, see http://publish.aps.org/PACS/pacs99.html 
 %Supernovae  
\bigskip  
]\renewcommand{\thefootnote}{\arabic{footnote}} \setcounter{footnote}{0}
\narrowtext  
%\widetext  

The observed relation between luminosity distance and redshift for
extragalactic Type~Ia Supernovae (SNe) appears to favor an
accelerating Universe, where almost two-thirds of the critical energy
density may be in the form of a component with negative pressure
\cite{perlmutter99,riess98,white1,garnavich,white2}.  Although this is
consistent with $\omegam<1$ and a cosmological constant $\Lambda>0$
(\eg\ \cite{apm}), at the theoretical level a constant
$\Lambda$ runs into serious difficulties, since the present value of
$\Lambda$ is $\sim$10$^{123}$ times smaller than predicted by most
particle physics models\cite{lambdarev}.

However, neither the present data nor the theoretical models require
$\Lambda$ to be exactly constant.  To explore the possibility that the
$\Lambda$-like term (\eg\ quintessence) is time-dependent, we use a
model for it that mimics the simplest variant of the inflationary
scenario of the early Universe.  A variable $\Lambda$-term is
described in terms of an effective scalar field (referred to here as
the $\Lambda$-field) with some self-interaction $V(\phi)$, which is
minimally coupled to the gravitational field and has little or no
coupling to other known physical fields. In analogy to the
inflationary scenario, more fundamental theories like supergravity or
the M-theory can provide a number of possible candidates for the
$\Lambda$-field but do not uniquely predict its potential
$V(\phi)$. On the other hand, it is remarkable that $V(\phi)$ may be
directly reconstructed from present-day cosmological observations.

The aim of the present letter is to go from observations to
theory, \ie\ from $D_L(z)$ to $V(\phi)$, following the prescription
outlined by Starobinsky \cite{starob98} 
(see also \cite{hut-tur}). 
This is the first attempt at reconstructing $V(\phi)$ from real 
observational data without resorting to
specific models (\eg\ cosmological constant, quintessence etc.). 

Since the spatially flat Universe ($\omegal\!+\!\omegam\!=\!1$) is both
predicted by the simplest inflationary models and agrees well with
observational evidence, we will not consider spatially curved
Friedmann-Robertson-Walker (FRW) cosmological models.  In a flat FRW
cosmology, the luminosity distance $D_L$ and the coordinate distance
$r$ to an object at redshift $z$ are simply related as ($c = 1$ here
and elsewhere)
\begin{equation}
a_0r = a_0 \int_t^{t_0} \frac{dt'}{a(t')} =\frac{D_L(z)}{1+z}.
\label{eqn:lumdis}
\end{equation}
This uniquely defines the Hubble parameter
\begin{equation}
H(z)\equiv \frac{\dot{a}}{a}
=\left[ \frac{d}{dz} \left( \frac{D_L(z)}{1+z} \right) \right]^{-1}.
\label{eqn:hfz}
\end{equation}
Note that this relation is purely kinematic and depends neither upon
a microscopic model of matter, including a $\Lambda$-term,
nor on a dynamical theory of gravity.

For a sample of objects (in this case, extragalactic SNe~Ia) for which
luminosity distances $D_L$ are measured, one can fit an
analytical form to $D_L$ as a function of $z$, and then estimate
$H(z)$ from (\ref{eqn:hfz}). 
If $\rho_m=(3H_0^2/8\pi G)\omegam(a/a_0)^{-3}$ 
is the density
of dust-like cold dark matter and the usual baryonic
matter, then 
{\setlength\arraycolsep{2pt}
\begin{eqnarray}
H^2 & = & \frac{8}{3}\pi G \left(\rho_m + {1\over 2}\dot\phi^2+ 
      V(\phi)\right),
                        \label{eqn:hsq}
\end{eqnarray}}
from where it follows that
\begin{equation}
\dot H=-4\pi\, G(\rho_m + \dot\phi^2). \label{eqn:alphaeq}
\end{equation}
Eqs.~(\ref{eqn:hsq}) \& (\ref{eqn:alphaeq}) can be rephrased 
in the following form convenient for our current reconstruction
exercise,
{\setlength\arraycolsep{2pt}
\begin{eqnarray}
{8\pi G\over 3H_0^2} V(x)\ &=& {H^2\over H_0^2} 
-{x\over 6H_0^2}{dH^2\over dx} -{1\over 2}\omegam\,x^3,
\label{eqn:Vzed}\\
{8\pi G\over 3H_0^2}\left({d\phi\over dx}\right)^2 &=& 
       {2\over 3H_0^2 x}{d\ln H\over dx} 
  -{\omegam x\over H^2},\label{eqn:phidot}
\end{eqnarray}}
where $x\equiv 1+z$.
Thus from the luminosity distance $D_L$, both $H(z)$ and $dH(z)/dz$
can be 
unambiguously calculated. 
This allows us to reconstruct the potential
$V(z)$ and $d\phi/dz$ if the value of $\Omega_M$ is 
additionally given.
Integrating the latter equation,
we can determine $\phi(z)$ (to within an additive constant) 
and, therefore, reconstruct the form of $V(\phi)$. Note also that the
present Hubble constant $H_0\equiv H(z\!=\!0)$ enters in a multiplicative
way in all expressions. Thus, neither the potential
$V(\phi)/H_0^2$ nor the cosmic equation of state $w_\phi(z)$
depends upon the actual value of $H_0$.

\noindent
\begin{figure}[thb!]
\centerline{  
%figure=dlfit.ps
\psfig{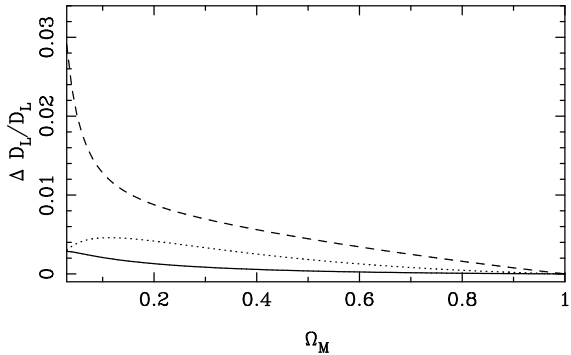}}
\medskip
{\caption{The maximum deviation $\Delta D_L/D_L$
between the actual value and that calculated from
the ansatz (\protect\ref{eqn:star})
in the redshift range $z=$0--10,
as a function of $\omegam\equiv 1-\omegal$. The three curves
plotted are for constant values of the equation of state parameter
(as defined in Eq.~{\protect\ref{eqn:wzed}})
$w_\phi=-1$ (solid line), $-2/3$ (dotted line) and $-1/3$ (dashed line).
}
\label{fig:ansatzplot}}
\end{figure}

\medskip

\noindent
{\it A fitting function for $D_L$:}
We use a rational (in terms of $\sqrt x$) {\it ansatz} for the
luminosity distance $D_L$, 
\begin{equation}               
{D_L\over x} 
\equiv \frac{2}{H_0}\left[ \frac{x - \alpha\sqrt{x} -1 + \alpha}{\beta x+ 
\gamma\sqrt{x} + 2 - \alpha -\beta -\gamma}\right]      
\label{eqn:star}  
\end{equation}
where $\alpha$, $\beta$ and $\gamma$ 
are fitting parameters. This function has the 
following important
features: it is valid for a wide range of models, and it is {\it
exactly equal} to the analytical form given by (\ref{eqn:lumdis}) for
the two extreme cases: $\omegal=0,1$. At these two limits, 
as $\omegam\to 1$,
$\alpha + \gamma \to 1$ and $\beta\to 1$ ; and as
$\omegam\to 0$,
$\alpha, \beta, \gamma \to 0$. 
The accuracy of our ansatz is illustrated in 
Fig.~\ref{fig:ansatzplot}.

We choose this form since the value of $H(z)$ obtained by
differentiating $D_L/x$, according to ({\ref{eqn:hfz}}), has the
correct asymptotic behavior: $H(z)/H_0 \to 1$ as $z\to 0$,
and  $H(z)/H_0 = \tilde\omegam^{1/2} (1+z)^{3/2}$ for $z\gg 1$, where
\begin{equation}
\tilde\omegam= \left({\beta^2\over \alpha \beta +\gamma}\right)^2.
\label{eqn:omegam}
\end{equation}
This ensures that at high-$z$, the Universe has gone through a matter
dominated phase.  It should be noted that $\tilde\omegam$ can be
slightly larger than the CDM component $\omegam$ since the
$\Lambda$-field (or quintessence) can have an equation of state
mimicking cold matter (dust) at high redshifts. For instance,
$\tilde\omegam \simeq 1.1\, \omegam$ in the quintessence model
considered by Sahni \& Wang \cite{sahni_wang}.  On the other hand,
$\tilde\omegam \simle 1.15\, \omegam$, to ensure that there is
sufficient growth of perturbations during the
matter-dominated epoch (see, \eg, the relevant discussion in
\cite{starob98}).

Note that the right hand side of ({\ref{eqn:phidot}}) should be
non-negative for the  minimally coupled scalar field model. At
$z=0$, this condition gives
\begin{equation}
{{4\beta +2\gamma -\alpha}\over {2-\alpha}}\ge 3\omegam,
\label{eqn:constraint1}  
\end{equation}
where the equality sign occurs when the $\Lambda$-term is constant.
The fact that $D_L$ is smaller in  a 
universe with time-dependent $\Lambda$-term than 
it is in a constant-$\Lambda $ universe leads to a lower 
limit for the parameter
$\beta$. When taken together with the fact that $\beta \to 1$
as $ \omegam \to 1$ ($\omegal \to 0$)
this leads to the following set of constraints
\begin{equation}
1 \le \frac{1}{\beta} \le  
\frac{1}{2}\int_1^{\infty}{dx\over \sqrt{1-\omegam+\omegam x^3}}.
\label{eqn:constraint2}  
\end{equation}

\noindent
\begin{figure}[thb!]
\centerline{  
%hubmilne.ps
\psfig{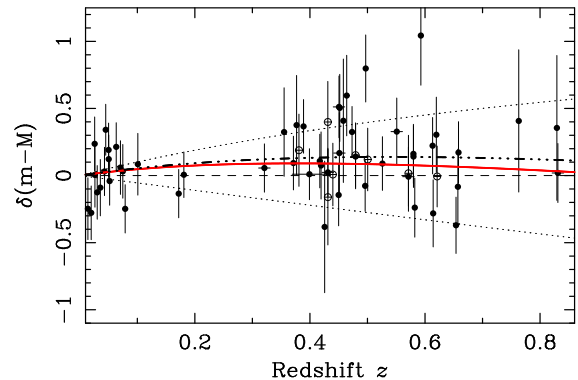}}
\medskip
{\caption{
The distance modulus $(m-M)$ of the SNe~Ia relative to an $\omegam\to 0$
Milne Universe (dashed line),
together with the
best-fit model of our ansatz (\protect\ref{eqn:star}), 
plotted as the solid line. 
The extreme cases of the  ($\omegam$, $\omegal$)= (0, 1)
and (1, 0) universes are plotted as dotted lines.
Also plotted as the dot-dashed line is  
the best fit Perlmutter \etal\ {\protect\cite{perlmutter99}}
model ($\omegam$, $\omegal$)= (0.28, 0.72).
The filled circles are the 54 SNe of 
the ``primary fit'' of {\protect\cite{perlmutter99}}. 
The high-$z$ SNe of 
{\protect\cite{riess98}} (not used in this analysis)
are plotted as open circles.  
}
\label{fig:hubbleplot}}
\end{figure}

\smallskip

\noindent
{\it The observational data:} Till date, about 100 SNe~Ia in the
redshift range $z\!=0.1\!-\!1$ have been discovered, a large fraction
of which have reliable published data from which luminosity distances
can be calculated.  We use the 54 SNe~Ia from the preferred ``primary
fit'' (`C' in their Table~1) of the Supernova Cosmology Project
\cite{perlmutter99}, including the low-$z$ Calan Tololo sample
\cite{hamuy96} as used therein.  We adopt the quoted
redshifts, reducing them to the cosmic microwave background frame. 

\begin{table}[bth]
\caption{Best-fit parameters
\label{table1}}
\begin{tabular}{lcrrrr}
$\omegam=$ && 0.2 &0.25  &0.3& \\
\tableline
$\alpha=$&&   $1.702^{+0.191}_{-0.518}$ &  $1.630^{+0.209}_{-0.539}$ 
                                         &  $1.533^{+0.229}_{-0.575}$ \\
$\beta=$&&   $0.585^{+0.019}_{-0.020}$ 
       &$0.604^{+0.022}_{-0.023}$   &$0.615^{+0.025}_{-0.027}$ \\
$\gamma=$&&  $-0.230^{+0.043}_{-0.045}$
                                & $-0.255^{+0.048}_{-0.052}$ 
                                   &$-0.252^{+0.056}_{-0.061}$ \\
$\kappa=$&&   $1.226\pm 0.018$ 
       &$1.229\pm 0.018$   &$1.232\pm 0.018$ \\
$\langle\chi^2\rangle$=&&1.03 & 1.03 & 1.03
&\tablenote{$\langle\chi^2\rangle=\chi^2_{\rm min}/(N-6)$, 
where $N=54$. 
The fit of four variables is subject to the constraint
$\tilde\omegam=\omegam$, and the inequality
(\ref{eqn:constraint1}), which significantly restricts the
permitted region.} \\
%so that we can use it as an additional constraint.} \\
$\tilde\omegam$=&&$0.2\pm 0.11$&$0.25\pm 0.16$&$0.3\pm 0.23$
\end{tabular}
\end{table}

\smallskip

\noindent
{\it 
Maximum likelihood fits:}
The luminosity distance $D_L$ (Mpc) is related to the measured
quantity, the corrected apparent peak $B$ magnitude $m_B$ as 
$ m_B = M_0 + 25 + 5\,\log_{10} D_L$,
where $M_0$ is the absolute peak luminosity of 
the SN. The function to be minimized is
\begin{equation}     
\chi^2 \equiv 
\sum_{i=1}^{n} 
\frac {\left[y(z_i)- y(m_{Bi})\right]^2}{\sigma_i^2};\ ~
y(z) \equiv 10^{M_0/5} D_L(z).
\end{equation} 
A fourth fitting parameter, $\kappa= 2\times 10^{M_0/5}(c/H_0)$,
which is required in addition to 
$\alpha,\beta,\gamma$ in the above minimization process,
includes both $M_0$ and $H_0$, which cannot be measured
independent of each other. For instance, if $M_0\!=\! -19.5\pm 0.1$
and $\omegam\!=\!0.3$, the value of $H_0\!=\!61.3\pm 2.9$ \kmsmpc.
Note that $\kappa$ only
features in the fit of (\ref{eqn:star}) to the data, and does not play a
role in the reconstruction of $V(\phi)$.

To obtain the best fit model, we perform an orthogonal chi-square fit, 
using errors on both the magnitude
and redshift axes in $\sigma_i$,
subject to the constraints (\ref{eqn:constraint1}), (\ref{eqn:constraint2})
and the condition $\tilde\omegam \simeq \omegam$.
The latter condition is used for simplicity -- our results remain
essentially the same even if we use the entire permitted range 
$\tilde\omegam \simle 1.15\, \omegam$.

%  Furthermore we require $ \alpha\! \le\! 2$ to ensure that the ansatz $
%  D_ L$ remains positive.  

The results shown in Table~{\ref{table1}} and in
Figure~{\ref{fig:hubbleplot}} are for $\omegam= 0.3$. In arriving at
the best fit, the two constraints in (\ref{eqn:constraint2}) are found
to be redundant, which means that only two constraints,
(\ref{eqn:constraint1}) and $\tilde\omegam =\omegam$, are actually
used.

\begin{figure}[bth!]
\centerline{  
%vzed.ps
\psfig{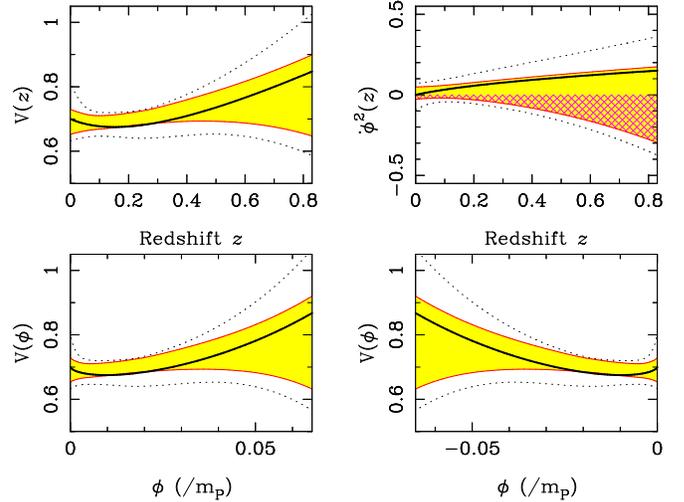} }
\medskip
{\caption{ The effective potential $V(z)$, and the kinetic energy term
${\dot\phi}^2$, are shown in units of $\rho_{\rm cr}=3H_0^2/8\pi G $.
Also plotted are the two forms of $V(\phi)$ for this $V(z)$, where the
errors do not reflect errors in the $z$-$\phi$ relation.  The value of
$\phi$ (known up to an additive constant) is plotted in units of the
Planck mass $m_{\rm P}$.  The solid line corresponds to the best-fit
values of the parameters.  In each case, the shaded area covers the
range of 68\% errors, and the dotted lines the range of 90\% errors.
The hatched area represents the unphysical
region ${\dot\phi}^2\!< \! 0$.}
\label{fig:vzed}}
\end{figure}

\smallskip
 
\noindent
{\it Reconstructing the scalar field potential:} We show the form of
the effective potential $V(z)$ reconstructed using ({\ref{eqn:Vzed}})
in Fig.~\ref{fig:vzed}, along with the corresponding plot for
$V(\phi)$, where $\phi$ is calculated by integrating
({\ref{eqn:phidot}}). The field $\phi$ is determined up to an
additive constant $\phi_0$, so we take $\phi$ to be zero at the
present epoch ($z=0$).

Our experiments with several realizations of synthetic data show that
this method works best if we fix the value of $\omegam$. Henceforth,
all reconstructed quantities are shown for $\omegam=0.3$.

For a scalar field, the pressure
$p\equiv -T^\alpha_\alpha={1\over 2}\dot{\phi}^2 - V$ and the energy density
$\varepsilon \equiv T^0_0={1\over2}\dot{\phi}^2 + V$ are related by
the equation of state,
\begin{eqnarray}
w_\phi (x) \equiv {p\over\varepsilon} &=& 
\frac{(2x/3) d\ln H/dx -1}{1-\left(H^2_0/H^2\right) 
\omegam x^3}.\label{eqn:wzed}
\end{eqnarray}
For the Cosmological constant, $w\!=\! -1$, while
quintessence models \cite{quintref} generally require $-1\!\le\!
w\!\le\! 0$ for $z\!\simle\!2$.

\begin{figure}[tbh!]
\centerline{  
%wzed.ps
\psfig{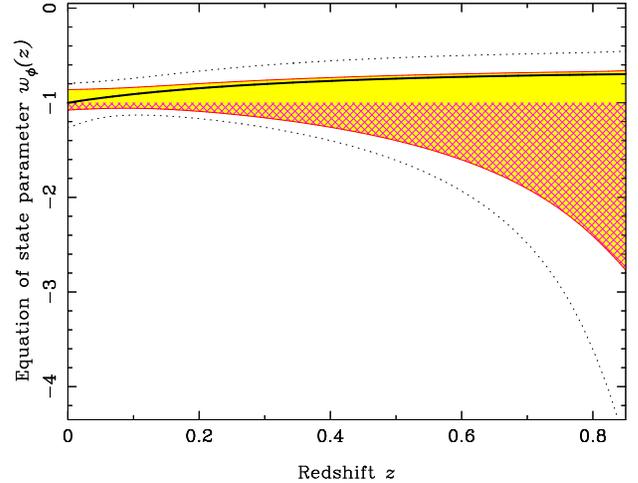} }
\medskip
{\caption{ The equation of state parameter $w_\phi(z)\!=\! P/\rho$ as
a function of redshift.  The solid line corresponds to the best-fit
values of the parameters.  The shaded area covers the range of 68\%
errors, and the dotted lines the range of 90\% errors.
The hatched area
represents the region $w_\phi\!\le\! -1$, which is disallowed for a
minimally coupled scalar field.}
\label{fig:wzed}}
\end{figure}

Our reconstruction for $w_\phi(z)$ according to (\ref{eqn:wzed}) is
plotted in Fig.~\ref{fig:wzed}. 
There is some evidence of possible evolution in $w_\phi$
with $-1\!\le\! w_\phi\!\simle\! -0.86$ preferred at the present epoch, and
$-1\!\le\! w_\phi \!\simle\! -0.66$ at $z=0.83$, the farthest SN in the
sample (both at 68\% confidence, upper limits correspond to
$-0.80$ and $-0.46$ at 90\% confidence respectively). 
However, a cosmological {\it constant} with $w=-1$ is
consistent with the data.

The errors quoted in this paper are calculated using a Monte-Carlo
method, where, in a region around the best-fit values of the
parameters shown in Table~1, random points are chosen in parameter
space from the probability distribution function given by the
$\chi^2$-function that is minimized to yield the best fit. At each
value of $z$ in the given range, the function in question is evaluated
at over $10^7$ such points, and the errors enclosing 68\% and 90\% of all
the values centered on the median are shown in the figures.
%(As an experiment, we also included an extra term in the $\chi^2$ function
%representing the variation of $\omegam$ as a Gaussian random variate
%with mean $\omegam=0.3$ and $\sigma=0.05$, but it made no significant change
%in the errors shown in these figures.)

\smallskip

\begin{figure}[tbh!]
\centerline{
%aged.ps  
\psfig{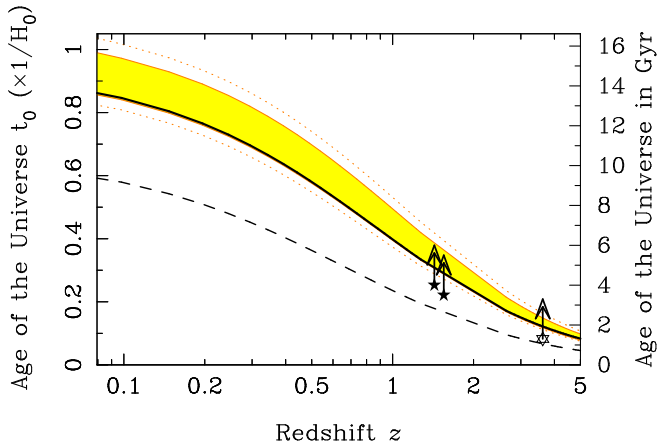} }
\medskip
{\caption{ The age of the Universe at a redshift $z$, given in units
of $H_0^{-1}$ (left vertical axis) and in Gyr, for the value of
$H_0\!=\!61.3$ \kmsmpc\ (right vertical
axis). The shaded region represents the range of 68\% errors, and
the dotted lines the range of 90\% errors.
The three high-redshift objects for which age-dating has
been published {\protect\cite{dunlop-yoshii}} are plotted as lower
limits to the age of the Universe at the corresponding redshifts. The
dashed curve shows the same relation for an ($\omegam, \omegal$)=(1,0)
Universe for the same $H_0$.  }
\label{fig:agefig}}
\end{figure}

\noindent
{\it The ages of objects:}
Our ansatz (\ref{eqn:star}) 
also provides us with a model-independent means of finding the age
of the universe at a redshift $z$,
\begin{equation}
t(z) = H_0^{-1}\int_z^\infty \frac{dz^\prime}{(1 + z^\prime)h(z^\prime)},
\label{eqn:age7}
\end{equation}
where the value of $h(z)\equiv H(z)/H_0$ is determined from 
({\ref{eqn:hfz}}). Figure~\ref{fig:agefig} 
shows the age of the Universe at a given $z$ and compares it with the
ages of two high redshift galaxies and the quasar
B1422+231 \cite{dunlop-yoshii}. We find that the requirement that the
Universe be older than any of its constituents at a given redshift
is consistent with our best-fit model, 
which is a positive feature since a flat matter-dominated
Universe must have an uncomfortably small value of $H_0$ to achieve this.

\noindent

\smallskip

\noindent
{\it Discussion:} In this letter, we have proposed a simple,
analytical, three parameter ansatz describing the luminosity distance
as a function of redshift in a flat FRW universe. The form of this
ansatz is very flexible and can be applied to determine $D_L$ either
from supernovae observations (as we have done) or from other
cosmological tests such as lensing, the angular size-redshift
relation etc. Using the resulting form of $D_L$ we reconstruct the
potential of a minimally coupled scalar $\Lambda$-field (or
quintessence) and its equation of state $w_\phi(z)$.  It should be
noted that the basic equations of this ansatz: (\ref{eqn:hfz}),
(\ref{eqn:star}), (\ref{eqn:wzed}) \& (\ref{eqn:age7}) are flexible
and can be applied to models other than those considered in the
present paper. For instance one can venture beyond minimally coupled
scalar fields by dropping either one or both of the constraints
(\ref{eqn:constraint1}) \& (\ref{eqn:constraint2}) (this is equivalent
to removing the constraint $\rho_\Lambda+p_\Lambda\! \geq\! 0$ on
the $\Lambda$-field).  Even with the limited
high-$z$ data currently available, our ansatz gives interesting
results both for the form of $V(\phi)$ as well as $w_\phi(z)$.  As
data improve, our reconstruction promises to recover `true' {\em
model-independent} values of $V(\phi)$ and $w_\phi(z)$ with
unprecedented accuracy, thereby providing us with a deep insight into
the nature of dark matter driving the acceleration of the universe.

\smallskip
\noindent{\it Acknowledgments:}
TDS thanks the UGC for providing support
for this work. VS acknowledges support from the
ILTP program of cooperation between India and Russia.  AS was
partially supported by the Russian Foundation for Basic Research,
grant 99-02-16224, and by the Russian Research Project
``Cosmomicrophysics''.


\begin{references}
%\setlength{\parskip}{0.3ex plus 0.2ex minus 0.5ex}
\bibitem[*]{byline} E-mail: saini@iucaa.ernet.in, 
somak@iucaa.ernet.in, 
varun@iucaa.ernet.in; alstar@hammer.landau.ac.ru
\bibitem{perlmutter99} S.J.~Perlmutter \etal, \apj, {\bf 517}, 565 (1999).
\bibitem{riess98} A.~Riess \etal, \aj, {\bf 116}, 1009 (1998). 
\bibitem{white1} M. White, \apj, {\bf 506}, 495 (1998).
\bibitem{garnavich} P.M. Garnavich \etal, \apj, {\bf 509}, 74 (1998).
\bibitem{white2} S.J. Perlmutter, M.S. Turner, and M. White, \prl, {\bf 83},
  670 (1999)
\bibitem{apm} G. Efstathiou, W. Sutherland, \& S. Maddox,
      \nat, {\bf 348}, 750 (1990)
\bibitem{lambdarev} V. Sahni, \& A. A. Starobinsky, IJMP, to appear (2000);
     also astro-ph/9904398
\bibitem{starob98} A.~A. Starobinsky, JETP Lett., {\bf 68}, 757 (1998).
\bibitem{hut-tur} D. Huterer, \& M. S. Turner, \prd, {\bf 60} 81301 
     (1999); T. Nakamura, \& T. Chiba, \mn, {\bf 306}, 696 (1999).
\bibitem{sahni_wang} V. Sahni \& L. Wang,  astro-ph/9910097.
\bibitem{hamuy96} M.~Hamuy \etal, \aj, {\bf 112}, 2391 (1996).  
\bibitem{quintref} R. R. Caldwell, R. Dave, \& P.~J.~Steinhardt,
 \prl, {\bf 80}, 1582 (1998);
N. A. Bahcall \etal,
          {\it Science}, {\bf 284}, 1481 (1999); 
   L. Wang, R.R. Caldwell, J.P. Ostriker, \& P.J. Steinhardt, 
   \apj, {\bf 530}, 17 (2000).
\bibitem{dunlop-yoshii} J. Dunlop \etal, \nat, {\bf 381}, 581 (1996); 
Y. Yoshii, T. Tsujimoto, \& K. Kawara, \apj,  {\bf 507}, L113 (1998)  
\end{references}
\end{document}